\documentclass[12pt]{article}
\usepackage{graphicx}

\newcommand{\bea}{\begin{eqnarray}}
\newcommand{\eea}{\end{eqnarray}}
\newcommand{\be}{\begin{equation}}
\newcommand{\ee}{\end{equation}}
\newcommand{\nn}{\nonumber}

\begin{document}
%
%
\begin{center}
\title{Neutrino Masses and Flavour Symmetry}
\author{\large G. Costa and E. Lunghi \\
Dipartimento di Fisica Galileo ``Galilei", Universit\`a di Padova \\
Istituto Nazionale di Fisica Nucleare, Sezione di Padova} 
\date{}
\maketitle
\end{center}
%
%
%
%
\begin{abstract}
The problem of neutrino masses and mixing angles is analysed
in a class of supersymmetric grand unified models, with
$SO(10)$ gauge symmetry and global $U(2)$ flavour symmetry.
Adopting the seesaw mechanism for the generation of the
neutrino masses, one obtains a mass matrix for the left--handed
neutrinos which is directly related to the parameters of the
charged sector, while the unknown parameters of the right--handed
Majorana mass matrix are inglobed in a single factor. 
\end{abstract}
\begin{center}
\begin{tabular}{|lll|} \hline
\bf PACS 12.10 & \vphantom{\fbox{$\bigcap$}}\bf -- & \bf Unified Field Theories and Models. \\
\bf PACS 12.15.Ff & \bf -- & \bf Quark and Lepton Masses and Mixing. \\ \hline
\end{tabular}
\end{center}
%
%
%
%
\section{Introduction}
The problem of neutrino masses is of great importance both for 
particle physics and astrophysics. However, it is still an
open problem: on one hand, the experimental indications 
of neutrino oscillations would indicate non-vanishing
neutrino masses and mixing angles; on the other hand, the
predictions for such quantities are strongly dependent on the
specific theoretical assumptions. Some of the main questions
of neutrino physics should 
be resolved by the new generations
of oscillation experiments which are under way (for a review see 
ref.[1]), and then one would be able to discriminate among the
different theoretical schemes. \par
While the minimal version of the Standard Model (SM) accomodates
only massless neutrinos, several different extensions have
been proposed, in which neutrino masses and mixing are
generated. In the present note, we shall adopt the simplest
form of the seesaw mechanism [2], in which right-handed
neutrinos are introduced and both Dirac $M_D\equiv M_{LR}$ and
    Majorana  $M_M\equiv M_{RR}$ mass matrices are generated.
As a consequence, the light neutrino masses are of the
type $M_{\nu}=M_D M_{M}^{-1} M_D^T$. While $M_D$ can be connected
in grand unified schemes to the charged lepton masses,
the Majorana mass matrix $M_M$ appears to be decoupled, in general,
from the charged sector. \par
Many theoretical investigations have been devoted to the analysis
of the mass spectrum of quarks and charged leptons. Most of them are
based on supersymmetric grand unified models, thus keeping the nice
features of gauge unification [3] and of the equality of the
b-quark and $\tau$-lepton masses at the unification scale.
The pattern of quark and lepton masses is characterized by a
strong hierarchy, which in the case of down-type quarks can be
approximately described by
%
\begin{equation}  m_d:m_s:m_b \approx \lambda^4:\lambda^2:1 ,
\end{equation}
where $\lambda \simeq 0.22$ is the Cabibbo angle, and by
%
\begin{equation}  m_u:m_c:m_t \approx \lambda^8:\lambda^4:1 ,
\end{equation}
for the up-type quarks. The case of the charged leptons is roughly
similar to (1). Such hierarchies lead to the ansatz of zero
textures, first employed by Fritzsch [4] and then analysed in
detail for the Yukawa matrices by Ramond, Roberts and Ross [5].
These textures suggested the introduction of an underlying ``family"
symmetry, and several schemes were proposed based on discrete
or continuous groups (we refer for details to the review papers [6]).
The Abelian group $U(1)$ (either global or local) has been one of the most 
favoured:
it can be interpreted as the remnant of a larger ``family"
or ``flavour" group $U(3) \subset U(45)$, which acts among the three
fermion families, or among all the different flavours. \par
An approximate flavour $U(2)$ symmetry has recently been proposed [7,8]
as an interesting framework for understanding the role of 
flavour breaking in supersymmetric theories. It is more useful
than the complete $U(3)$ symmetry, which is badly broken by the large
top-quark mass, and it is more predictive than the Abelian
$U(1)$ symmetry. While solving the supersymmetric flavour-changing
problem, it leads to interesting relations among the masses and
mixing angles, some of which appear to be well satisfied [9].
Specific models based on $SU(5) \otimes U(2)$ and $SO(10) \otimes
U(2)$ have been analysed [9,10]. \par
More recently, the
analysis based on the $U(2)$ flavour symmetry has been extended
to the neutrino sector [11], with some interesting results for the
neutrino phenomenology. The model, which is based on $SU(5)\otimes U(2)$
will be briefly discussed in the following; in alternative, we 
present here a class of models based on the symmetry $SO(10) \otimes U(2)$
which exhibit an important
advantage. In fact, one can show that the light neutrino mass matrix
can be directly related to the parameters of the quark and charged lepton
mass matrix, while the unknown parameters of the right-handed neutrino
Majorana mass matrix are inglobed in a single factor. \par
In Section 2, we outline the general features 
of the $U(2)$ flavour symmetry models. In Section 3
we review what has been done in the frame of $SU(5)\otimes U(2)$
scheme, both for the charged fermions and for the neutrinos; 
in Section 4 we go to $SO(10) \otimes U(2)$ and describe our results
for the neutrino sector; finally, we draw our conclusions.
%
%
%
%
\section{Fermion masses in $U(2)$ flavour symmetry models}
While we refer to the review papers [6] for the information about the different
approaches adopted for the problem of the fermion
mass hierarchies, we outline here the theoretical framework
for the specific case of the $U(2)$ flavour symmetry [7,8]. \par
Remaining, for the moment, at the level of the Minimal
Supersymmetric Standard Model (MSSM), the internal symmetry group is
specified by:
%
\begin{equation}    
G^{int} = G^{int}_{SM} \otimes U(2). \end{equation}
We denote by $\psi$ all the chiral superfields which include the whole 
fermion sector 
of the SM, and by $h$ the two Higgs doublets $h_u$ and $h_d$; they
are assigned to the $U(2)$ representations as follows:
$\psi = \psi_a + \psi_3 = {\bf 2} + {\bf 1} \; (a=1,2)$ and
$h = {\bf 1}$. \par
The Yukawa interactions can be obtained from an operator expansion of 
the superpotential in 
an effective theory, in terms of a set of fields $f$, called ``flavons",
which develop vacuum expectation values (vevs) breaking the $U(2)$ symmetry:
%
\begin{equation}   W = {[\psi h (1 + f/M + f^2/{M^2} + ...) \psi]}_F .
\end{equation} 
and $M$ is the cut-off of the effective theory. \\
The flavons can be assigned only to the ${\bf 2}$ and ${\bf 2}\times 
{\bf 2} = {\bf 3}_s + {\bf 1}_a$ representations of $U(2)$
($\phi^a = {\bf 2}, S^{ab} = {\bf 3}_s, A^{ab} = 
{\bf 1}_a$) and the effective superpotential becomes to order $f/M$:
%
\begin{equation}  W = \psi_3 h \psi_3 + {1 \over M} \psi_3 h \phi^a \psi_a
+ {1 \over M} \psi_a h (S^{ab} + A^{ab}) \psi_b. \end{equation}
It is assumed that 
the symmetry $U(2)$ is broken spontaneously in two steps by the 
vevs $\langle \phi^2 \rangle$, $\langle S^{22} \rangle$ and $\langle A^{12} 
\rangle$:
%
\begin{equation}  U(2) \buildrel \epsilon \over \longrightarrow U(1)
\buildrel \epsilon' \over \longrightarrow \O.
\end{equation} \par
The Yukawa matrices ($\lambda^{U,D}$ for the up--, down--type quarks and 
$\lambda^E$ for the charged leptons) are given by the same expression:
%
\begin{equation}  \lambda \sim \left( \begin{array}{ccc} 0 & 
\epsilon' & 0 \\ -\epsilon' & \epsilon & \epsilon \\ 0 & \epsilon & 1
\end{array} \right) \end{equation}
where $\epsilon \simeq \epsilon_S \simeq \epsilon_{\phi} \ll 1$
($\epsilon_S = \langle S^{22} \rangle /M$ , $\epsilon_{\phi}=
\langle \phi^2 \rangle /M$), and
$\epsilon' = \langle A^{12} \rangle /M \ll \epsilon$.
This structure is satisfactory from the phenomenological point of
view if [9]:
%
\begin{equation} \epsilon \approx m_s/m_b 
\end{equation}
\begin{equation}  \epsilon' \approx \sqrt{\frac{m_d m_s}{{m_b}^2}}.
\end{equation} 
However, there are some difficulties related to the fact that
the hierarchy in the up-like quark sector is
stroger than in the down-like quark and charged lepton sectors,
which can be exemplified by:
%
\begin{equation}  m_{\tau} \approx m_b \ll m_t . \end{equation}
These difficulties could be overcomed by discriminating between
$\lambda^U$ and $\lambda^D \approx \lambda^E$, and by imposing that 
the $\lambda^U_{22}$ and $\lambda^U_{21}$ entries vanish at order
$\epsilon$ and $\epsilon'$. These features can be realized in grand 
unified models, as will be shown in the next sections. \par
We note that the problem of FCNC, which arises in 
general in supersymmetric theories due to the presence of soft 
breaking terms, can find a natural solution in the $U(2)$ flavour
symmetry model [7]. The suppression of FCNC requires, in fact, mass
degeneracy of the sfermions of the first two families: in the $U(2)$
model the splitting is indeed very small (of order $\epsilon^2$).
On the other hand, the splitting between the third and the two
lighter families of sfermions could give rise to appreciable
contributions for observables like the $\mu \rightarrow e + \gamma$
decay and the $CP$-violating parameter $\epsilon_K$ and this
would be an interesting signature of the model [7].
%
%
%
%
\section{The $SU(5) \otimes U(2)$ model}
A way of suppressing the $\lambda^U_{22}$ and $\lambda^U_{21}$ entries,
keeping at the same time the corresponding $\lambda^{D,E}_{22}$ and 
$\lambda^{D,E}_{21}$ different from zero, can be realized in the frame of 
the $SU(5)$ grand unification. \par
With the usual assignements of a fermion family to the superfield $T_i (
{\bf 10})
+ \bar F_i ({\bf \bar 5})$, and of the two Higgs doublets to the superfields 
$H_u ({\bf 5})$ and $H_d ({\bf \bar 5})$, the superpotential (5) 
is replaced by:
%
\bea 
W_{(5)} & = & T_3 H_u T_3 + T_3 H_d \bar F_3 + \nn \\
        &   & {1\over M} (T_3 H_u \phi^a T_a + T_3 H_d \phi^a \bar F_a + 
\bar F_3 \phi^a H_d T_a) + \nn \\
        &   & {1\over M} [T_a (S^{ab} + A^{ab}) H_u T_b + T_a (S^{ab} + 
A^{ab}) H_d \bar F_b]. 
\eea
Here and in the following, an arbitrary complex constant of order $O(1)$ is
implied for each term of the superpotential. The fields $\phi^a$ can be 
assigned either to $\bf 1$ or to $\bf 24$ and a convenient 
choice for the other flavons is
%
\be S^{ab} = {\bf 75} \;\;\;\;\; \hbox{and} \;\;\;\;\; A^{ab} = {\bf 1}. \ee
In fact, as shown in [9], the above assignement, which implies that the 
terms $S^{ab} H_u$ and $A^{ab} H_u$ transform, respectively, as 
${\bf 45}$ and ${\bf 5}$, guarantees the vanishing of 
$\lambda^U_{22}$ and $\lambda^U_{21}$ at order $\epsilon$ and $\epsilon'$. \par
In order to obtain a non vanishing contribution for the $m_u$ mass at higher 
order, one has to introduce an extra familon 
$\Sigma_Y$ in the representation ${\bf 24} \otimes {\bf 1}$ of $SU(5) \otimes U(2)$, with a
vev in the direction of the hypercharge $Y$ with value
$\langle \Sigma_Y \rangle /M \simeq \rho \simeq 0.02$. \\
The additional terms in the superpotential are:
%
\be W'_{(5)} = {1\over M^2} T_a [\phi^a \phi^b + (S^{ab} + A^{ab}) 
\Sigma_Y] 
H_u T_b \ee
which generate the contributions $\lambda^U_{22} = O (\epsilon \rho)$, 
and $\lambda^U_{12} = O (\epsilon' \rho)$. \par
Finally, one can write for the Yukawa matrices
%
\bea \lambda^U & = & \lambda \pmatrix{0 & \epsilon' \rho & 0 \cr
- \epsilon' \rho & \epsilon \rho' & x_u \epsilon \cr
0 & y_u \epsilon & 1 \cr} \\
\lambda^{D,E} & = & \xi \pmatrix{0 & \epsilon' & 0 \cr
-\epsilon' & (1,-3) \epsilon & (x_d, x_e) \epsilon \cr
0 & (y_d, y_e)\epsilon & 1 \cr} \eea
where the complex coefficients $x_u,  x_d, x_e, y_u, y_d, y_e$ are of 
order 1, and $\rho' = O (1) \rho$. Moreover, to
account for the different hierarchies (1) and (2) one has to take $\xi << 
\lambda$, which  can be realized by assuming that the light Higgs (in the
unified multiplet $H_d$) which couples  to the D/E sector contains a 
small component (of order $m_b/m_t$) of the Higgs doublet [9]. \\
Disregarding the unknown factors of order 1, the matrices (14) and 
(15) depend only on 4 small parameters: $\rho \simeq \epsilon \simeq \xi 
\simeq 0.02$ and $\epsilon' \simeq 0.004$. \\
With this approximation one can relate the 13 observables of the fermion 
sector by means of 9 approximate relations; in fact, 5 of them are precise 
relations [9]. \par
So far neutrinos have been considered strictly massless. Let us now add a 
right--handed neutrino superfield $\nu_i$ to each family,  
singlet under $SU(5)$ and transforming as
%
\be \nu_i = (\nu_a, \nu_3) = ({\bf 2},{\bf 1}) \ee
under $U(2)$.\\
Correspondingly, the superpotential contains the additional terms:
%
\bea W^D_{(5)} &=& \bar F_3 H_u \nu_3 + {1\over M} [\bar F_3 \phi^a H_u \nu_a
+ \bar F_a \phi^a H_u \nu_3 + \bar F_a A^{ab} H_u \nu_b] \nn \\
               & & + {1\over M^2} \bar F_a (\phi^a \phi^b + S^{ab} \Sigma_Y) 
H_u \nu_b. \eea
They produce a Dirac mass matrix of the  form:
%
\be \lambda^\nu_D \sim \pmatrix{0 & \epsilon' & 0 \cr
-\epsilon' & \rho\epsilon & \epsilon \cr
0 & \epsilon & \rho \cr} \ee
where only the order of magnitude is indicated for each entry. \\
The Majorana mass matrix for the right--handed neutrinos is generated by 
the superpotential
%
\be W^M_{(5)} = \lambda_M \nu_3 \nu_3 + {\lambda_M \over M} [\nu_a \phi^a 
(1 + {\Sigma_Y 
\over M}) \nu_3 + \nu_a ({\phi^a \phi^b \over M}) 
\nu_b ]. \ee
We note that  terms containing $A^{ab}$ and $S^{ab}$ are not allowed by 
symmetry reasons and by the assignement (12). Since the superfield 
$\phi^a$ can be  assigned either to ${\bf 1}$
or ${\bf 24}$, two choices are possible for the 
Majorana mass  matrix:
%
\bea \lambda_M^{(1)} & \sim & \lambda_M \pmatrix{0 & 0 & 0 \cr
0 & \epsilon^2 & \epsilon \cr
0 & \epsilon & 1 \cr} \\
\lambda_M^{(24)} & \sim & \lambda_M \pmatrix{0 & 0 & 0 \cr
0 & \epsilon^2 & \rho\epsilon \cr
0 & \rho\epsilon & 0 \cr}. \eea 
Similar results are presented in ref. [11]. \\
The Majorana mass matrices obtained from (19) have zero determinant, 
thus preventing the use of the seesaw mechanism. A solution to this 
problem can be obtained with the introduction of  
other flavons in the representations $\bf 1$ or ${\bf 3}_s$ 
of $U(1)$, like $\phi^a$ or $S^{ab}$, but with non--vanishing vev's along 
the directions $\langle \phi^1 \rangle$ or $\langle S^{12} \rangle$. \par
However, this approach is rather dangerous
in $SU(5)$, since these flavons should be assigned  to the 
representation 1 or 24 of the gauge group,  and then they would couple to 
all the other fermion fields, spoiling the results  obtained  for the 
Yukawa matrices and for the  FCNC suppression. \\
In the quoted ref. [11], a solution with $\langle S^{11} \rangle =  
\langle S^{12} \rangle =0$ and $\langle \phi^1 \rangle 
\leq \epsilon'$ is 
considered to be compatible with  the present phenomenology of the quark 
sector. \par
In the next section, we propose an alternative  approach based on the
$SO(10) \otimes U(2)$ symmetry.
%
%
%
%
\section{Models based on $SO(10) \otimes U(2)$}
In the supersymmetric models based on the gauge group $SO(10)$ the chiral 
superfield $\psi_i$ contains all fundamental fermions, including the right 
handed neutrinos (16). The field $\psi_i$ and the Higgs fields transform 
under $SO(10) \otimes U (2)$ as follows:
%
\bea \psi_i &=& \psi_a + \psi_3 = {\bf 16} \otimes ({\bf 2} \oplus 
{\bf 1}) \; \; (a =1,2) \\
       H &=& h_u + h_d = {\bf 10} \otimes {\bf 1}. \eea
To first order in the familon fields, the superpotential becomes:
%
\be W_{(10)} = \psi_3 H \psi_3 + {1\over M} [\psi_3 \phi^a H \psi_a + 
\psi_a (S^{ab} + A^{ab  }) H \psi_b] \ee
where $\phi^a$, $S^{ab}$ and $A^{ab}$ belong to the $SO(10)$ 
representations {\bf 1}, {\bf 45}, {\bf 54} or 
{\bf 210}. \par
In the following, we consider in detail two different classes of models,
which were analysed in ref. [9] in connection with the charged fermion 
sector. Here we limit ourselves to specify only the main ingredients
referring to [9] for details. \par
\bigskip \noindent
%
%
%
%
{\bf I.} The first class of $SO(10)$ models consists in a direct extension of the
$SU(5)$ model described in  the previous sections. One makes the following 
correspondence between $SU(5)$ and $SO(10)$ flavons:
%
\bea S^{ab} ({\bf 75}) & \rightarrow & S^{ab} ({\bf 210}) \\
     A^{ab} ({\bf 1}) & \rightarrow & A^{ab} ({\bf 45}) \\
     \phi^a ({\bf 1}, {\bf 24}) & \rightarrow & \phi^a ({\bf 45}) \\
     \Sigma_Y ({\bf 24}) & \rightarrow & \Sigma_Y ({\bf 45}) 
\eea
and the vevs of the $SO(10)$ fields  have 
to be taken in the same directions of vev's of the 
corresponding $SU(5)$ fields. \par
With the superpotential (24), the entries $\lambda^U_{22}$ 
and $\lambda^U_{12}$ 
are suppressed to first order in $\langle f \rangle /M$, so 
that it is replaced by
%
\be W^I_{(10)} = W_{(10)} + W'_{(10)} \ee
where the additional terms 
%
\be W'_{(10)} = {1\over M^2} \psi_a [\phi^a \phi^b + (S^{ab} + A^{ab}) 
\Sigma_Y] H \psi_b \ee
are included, in analogy to what done in $SU(5)$, to produce higher order 
contributions to $m_u$. \par
The Yukawa matrices obtained from (29) have the same expression 
of (14) and (15); the only difference  being that the assignement of 
$\phi^a ({\bf 45})$ to a single representation reduce the number of the 
independent parameters
$x_a$ and $y_a \; (a = u, d, e)$.\\
In table 1 we show the order of magnitude obtained for the
charged fermion masses, and indicate the  flavons  which give 
non--vanishing contributions. \par
%
\begin{center}
\begin{tabular}{||c|c|c||} \hline
\bf mass & $\bf O()$ & \bf flavons \\ \hline
$ m_b, m_\tau$ & $1$ & $1$ \\
$ m_t$ & $1$ & $1$ \\
$m_s, m_\mu$ & $\epsilon$ & $\phi^a, S^{ab}$ \\ 
$m_c $ & $\rho\epsilon$ & $\phi^a, S^{ab} \sum_Y$ \\
$m_d, m_c$ & $\epsilon'$ & $A^{ab}$ \\
$m_u$ & $\rho\epsilon'$ & $A^{ab} \sum_Y$ \\ \hline
\end{tabular} \par \bigskip \noindent
{\bf \small Table 1}
\end{center}
In the present case, also the neutrino Dirac mass matrix is generated 
by the superpotential (29), which it is given by:
%
\be \lambda^{\nu I}_D = \lambda \pmatrix{0 & \epsilon' & 0 \cr
-\epsilon' & \rho' \epsilon & x_\nu \epsilon \cr
0 & y_\nu \epsilon & 1 \cr}. \ee \par
The situation is different for the right--handed Majorana mass matrix, since the 
product decomposition ${\bf 16} \otimes {\bf 16}$ does not contain 
the identity representation {\bf 1}. In fact, a term 
$\nu_R \nu_R$ would violate the $U(1)_X$ 
symmetry included in $SO(10)$. This term can be generated with the 
inclusion of flavons in the representation $\overline{\bf 126}$, 
and by assuming that their
vev's lies in the direction of the $SU(5)$ singlet with $X =-10$.\\
The following choices are possible:
%
\bea  
L & = & \overline{\bf 126} \otimes {\bf 1} \\
     L^a & = & \overline{\bf 126} \otimes {\bf 2} \\
     L^{ab} & = & \overline{\bf 126} \otimes {\bf 3}_s 
\eea
and their contributions to the superpotential are given by
%
\bea & & \psi_3 L \psi_3 + \psi_a L^a \psi_3 + \psi_a 
L^{ab} \psi_b \nn \\
    & & + {1\over M} [\psi_a \phi^a (1 + {\Sigma_Y \over M}) L \psi_3 + 
\psi_a (S^{ab} + A^{ab} + {\phi^a \phi^b \over M}) L \psi_b ]. \eea
Keeping only $L$ one gets Majorana mass matrices similar to those obtained in
the $SU(5)$ case, i.e. (20,21):
%
\bea \langle \Phi^a (45) \rangle & \sim & ({\bf 1}, 0) \longrightarrow \lambda_M 
\pmatrix{0 & 0 & 0 \cr
0 & \bar x \bar \epsilon^2 & \bar \epsilon \cr
0 & \bar \epsilon & 1 \cr} \\
\langle \Phi^a (45) \rangle & \sim & ({\bf 24},0) \longrightarrow \lambda_M 
\pmatrix{0 & 0 & 0 \cr
0 & \bar x \bar \epsilon^2 & \bar \rho \bar \epsilon \cr
0 & \bar \rho \bar\epsilon & 1 \cr} \eea
where (,) refers to the $SU(5) \otimes U(1)_X$ content and      
$\bar\epsilon \sim \epsilon$, $\bar\rho \sim \rho$, $\bar x \sim O(1)$.
The terms containing $L^a$ and $L^{ab}$ generate Majorana mass matrix of 
the  form
%
\be M_M = \lambda_M \pmatrix{\bar x_{11} & \bar x_{12} & \bar x_1 \cr
\bar x_{12} & \bar x_{22} & \bar x_2 \cr
\bar x_1 & \bar x_2 & 0 \cr} \ee
where we have used the notation: $\langle L^{ab} \rangle / \lambda_M 
= \bar x_{ab} \sim O(1)$  and $\langle L^a \rangle / \lambda_M 
= \bar x_a \sim O(1)$. \par
In order to constrain the Majorana mass matrix which, 
in general, would be the sum of
(38) and either (36) or (37), {\it we  
introduce the minimum set of $L$--type flavons, and assume that each them 
develops a single vev}. To avoid vanishing 
determinant, we need always two kinds of flavons, such as $(L, L^a)$, 
$(L, L^{ab})$ or $(L^a, L^{ab})$. Taking into account the two possibilities
(36) and (37), one can build 8 different kinds of matrices. \par
\bigskip
\noindent
%
%
%
%
{\bf II.} We consider here a second class of models 
[9] which are based on $SO(10)$ 
but that are not reducible to the $SU(5)$ models. The superpotential is 
given by (24) and the question related to 
the suppression of the $\lambda^U_{22}$ and $\lambda^U_{12}$ entries is 
solved, in the present case, making use of the specific structure of 
the $SO(10)$ representations. \par
The flavons are now assigned only to representations ${\bf 45}$ or 
${\bf 1}$:
%
\bea \phi^a & \longrightarrow & {\bf 45} \\
     S^{ab} & \longrightarrow & {\bf 45} \\
     A^{ab} & \longrightarrow & {\bf 1} \; \hbox{or} \; {\bf 45} \eea 
Different choices for the vev's of ${\bf 45}$ are possible, leading 
to alternative solutions of the problems.
\begin{description}
\item[$\;\;$ \bf a)] The vanishing  of $\lambda^U_{22}$ can be obtained  
by assuming $\langle S^{ab} \rangle = {\bf 45}_{B-L} $.
The operator $\psi_a S^{ab} H \psi_b$ in (24) vanishes identically because 
(B--L) has opposite value for isospin doublets  and singlets. In this case 
also $\lambda^{D,E}_{22}$ are suppressed. 
\item[$\;\;$ \bf b)] The vanishing  of $\lambda^U_{12}$ can be obtained with either $A^{ab} = 
{\bf 1}$ or $A^{ab} = {\bf 45}$. With the first 
choice the operator $\psi_a A^{ab} H \psi_b$ is suppressed by symmetry 
reasons, but in this case one gets also $\lambda^{D,E}_{12} =0$. With the 
second choice, the vev is taken along the $X$--direction; since $X(Q_L) = X 
(u^c_R)$, one gets $\lambda^U_{12} =0$ and $\lambda^{D,E}_{12} = 
O(\epsilon')$. 
\item[$\;\;$ \bf c)] Terms $\lambda^{D,E}_{22} = O(\epsilon)$ and $\lambda^{D,E}_{12} = 
O(\epsilon')$ can be obtained by introducing also a flavon $\Sigma_X 
({\bf 45})$  with vev  along the $X-$direction. The higher order contributions to 
$\lambda^U_{22}$ and $\lambda^U_{12}$ require the additional terms (30) 
with the inclusion also of $\Sigma_X$.
We do not reproduce here the terms of the superpotential which contain 
$\Sigma_X$, given explicity in ref. [9].
\end{description} \par
Also in this class of model the matrix $\lambda^U$ has the same form of (14);
there is some change of  sign in $\lambda^E$, and the explicit expression 
will be given later. \\
For the Dirac neutrino mass matrix, one obtains
%
\be  \lambda^{\nu II}_D = \lambda \pmatrix{0 & -2\epsilon'& 0 \cr
2\epsilon' & 6\epsilon & x_\nu \epsilon \cr                              
0 & y_\nu \epsilon & 1 \cr} \ee
where the parameters $x_\nu$ and $y_\nu$ depend on the direction of the 
vev of \hbox{$\phi^a (45)$}. \\
The analysis of the right handed Majorana mass matrix is similar to that 
performed for the models of the first class; 
the contribution of the $L$ field is now  given by
%
\be \lambda_M \pmatrix{0 & 0 & 0 \cr
0 & \bar y \bar \epsilon & \bar \epsilon \cr
0 & \bar \epsilon & 1 \cr} \ee
with $\bar y \sim O(1)$.\\
In the present case, with the inclusions of the possible pairs $(L, L^a)$, 
$(L, L^{ab})$ and $(L^a, L^{ab})$, one gets 5 different kinds of Majorana 
matrices. \par 
\bigskip 
Before discussing the specific properties of the different solutions for 
the neutrino masses, we collect the Yukawa matrices obtained for the quarks
and charged leptons in the two classes I and II:
%
\bea \lambda^U & = & \lambda \pmatrix{0 & \epsilon' \rho & 0 \cr
-\epsilon' \rho & \epsilon\rho' & x_u\epsilon \cr
0 & y_u\epsilon & 1 \cr} \\
\lambda^{(D,E)} & = & \xi \pmatrix{0 & (1, \pm 1) \epsilon' & 0 \cr
(1, \mp 1)\epsilon' & (1, \mp 3)\epsilon & (x_d, x_e)\epsilon \cr
0 & (y_d, y_e)\epsilon & 1 \cr} \eea
where the upper and lower signs in (45) refer to class I and II, 
respectively. It can be shown that the fermion masses and mixing depend on 
9 independent real parameters of the 24 introduced in (44, 45), so that 4 
precise predictions  among the 13 physical quantities are obtained [9]. The
independent parameters can be expressed in terms of the physical 
quantities; in particular one gets $\lambda \sim m_t, \xi \sim m_b \sim 
m_\tau,$ and  the two relations given in (8) and (9). \par
The neutrino Dirac mass matrices for the two classes I and II, are 
explicity:
%
\be m_D^{I,II} = m_t \pmatrix{ 0 & (1, -2)\epsilon' & 0 \cr
(-1, 2)\epsilon' & (\rho', 6)\epsilon & x_\nu \epsilon \cr              
0 & y_\nu \epsilon & 1 \cr} \ee
and the effective neutrino mass matrix becomes
%
\be M^{I,II}_\nu = m^{I,II}_D (M^{I,II}_{M})^{-1} (m_D^{I,II})^T. \ee
The quantities $x_\nu, y_\nu$ are model dependent, but they can be 
expressed in terms of $x_a$ and $y_a$ $(a = u, d, c)$,
so that the Dirac matrices are strictly linked to the charged fermion 
sector. On the other hand, the Majorana matrices depend on arbitrary 
parameters. As noted previously, there are different solutions for $M_M$ 
in both classes; however, in all cases one can write 
%
\be M_\nu^{(I,II)} = \langle m_\nu \rangle \; \lambda_\nu^{(I,II)} \ee
where $\langle m_\nu \rangle$ depend on $\lambda_M$ and possibly on other parameters of the 
Majorana matrix, while the $\lambda_\nu$ matrices determine the neutrino 
masses and mixing hierarchies. \par
It is important to point out that there are three solutions (one in the 
first class, and two in the second) which, aside from a common factor, 
do not depend on the parameters of the Majorana matrices. Leaving out a 
factor $O(1)$, the three solutions can be written as follows:
$$\begin{array}{lcl}
M^I_{M1} = \lambda_M \pmatrix{0 & 0 & \bar x_1 \cr
0 & \bar x \bar \epsilon^2 & \bar\rho \bar\epsilon \cr
\bar x_1 & \bar\rho \bar\epsilon & 1 \cr}
& \Rightarrow & 
M^I_{\nu 1}  \simeq  \displaystyle {m_t^2 \over \lambda_M} \pmatrix{\eta^2 &
\rho'\eta & y_\nu\eta \cr
\rho'\eta & \rho'^2 & y_\nu \rho' \cr
y_\nu \eta & y_\nu \rho' & y_\nu^2  \cr} \nonumber \\
&&\\
M^{II}_{M2} = \lambda_M \pmatrix{0 & 0 & \bar x_1 \cr
0 & \bar y \bar \epsilon & \bar \epsilon \cr
\bar x_1 & \bar\epsilon & 1 \cr}
& \Rightarrow & 
M^{II}_{\nu 2} \simeq \displaystyle {m^2_t \over \lambda_M} \pmatrix{4\eta & 
-12 \epsilon' & -2 y_\nu \epsilon' \cr
-12 \epsilon' & 36 \epsilon & 6 y_\nu \epsilon \cr
-2 y_\nu \epsilon' & 6 y_\nu \epsilon & y^2_\nu \cr} \nonumber \\
&&\\
M^{II}_{M3} = \lambda_M \pmatrix{\bar x_{11} & 0 & 0 \cr
0 & 0 & \bar x_2 \cr
0 & \bar x_2 & 0 \cr}
& \Rightarrow & 
M^{II}_{\nu 3} \simeq \displaystyle {m^2_t \over \lambda_M} \pmatrix{0 & -2
\epsilon\epsilon' x_\nu & -2\epsilon' \cr
-2 \epsilon\epsilon' x_\nu & 12\epsilon^2 x_\nu & 6\epsilon \cr
-2 \epsilon' & 6\epsilon & 2\epsilon y_\nu \cr} \nonumber
\end{array} $$
where $\eta = \epsilon'/\epsilon \approx \sqrt{m_d/m_s}.$ 
However, the first matrix has a zero eigenvalue; to obtain
massive eigenstates one should include also terms of order
$\epsilon^2$ in the Majorana mass matrix, but then new 
parameters, not determined by the charged fermion sector,
would be introduced, destroying the predictivity of the
model.

Finally, we are left with the two solutions $M^{II}_{\nu 2}$
and $M^{II}_{\nu 3}$. To test quantitatively these solutions, 
we should look for a specific realization of the superpotential 
adopted in this section. Alternatively, we can check if there
are regions in the parameter space, allowed by the constraints of the 
charged sector, for which the mass matrices $M^{II}_{\nu 2}$
and $M^{II}_{\nu 3}$ give results in agreement with the present
neutrino oscillation phenomenology. Specifically, we refer to the
three-flavour fit [13] of the solar neutrino deficit (assuming the 
Mikheyev-Smirnov-Wolfenstein oscillation mechanism) and the
atmospheric neutrino anomaly, disregarding the LSND experiment. 
In fact, while the first two pieces of information appear to be rather 
well established [1], the third one need further confirmation;
on the other hand, it is impossible to fit the three sets of data
in a model with no more than three massive neutrinos.

The mass spectrum and the mixing structure obtained from $M^{II}_{\nu 
2,3}$ are very sensitive to variations of the parameters $x_\nu$ and
$y_\nu$ (which characterize the different models), 
while they are relatively
invariant under those of $\epsilon$ and $\epsilon'$, for which 
we take the values $\epsilon=0.02$ and $\epsilon'=0.004$, in 
agreement with (8) and (9). A numerical
analysis shows that only for $M^{II}_{\nu 3}$ there are regions
of the paramenter space in which the experimental constraints
are satisfied. \par
Let us denote by $\nu_i \;\; (i = 1, \; 2, \; 3)$ the mass eigenstates
and by $\nu_\alpha \;\; (\alpha = e, \; \mu, \; \tau)$ the interaction
ones:
%
\be M^{II}_{\nu 3} \; \nu_i = m_{\nu_i} \; \nu_i \ee
\be \nu_\alpha = U_{\alpha i} \nu_i \ee
\be \left( U_{\alpha i} \right) =
\pmatrix{c_{e2}c_{e3} & s_{e2}c_{e3} &
s_{e3}e^{-i \delta}\cr -s_{e2}c_{\mu 3}-s_{e3}s_{\mu 3}c_{e2}e^{i\delta} &
c_{e2}c_{\mu 3}-s_{e2}s_{\mu 3}s_{e3}e^{i \delta} & s_{\mu 3}c_{e3}\cr
s_{e2}s_{\mu 3}-c_{e2}c_{\mu 3}s_{e3}e^{i \delta} &
-c_{e2}s_{\mu 3}-s_{e2}s_{e3}c_{\mu 3}e^{i \delta} & c_{\mu 3}c_{e3}\cr} \ee
where $m_{\nu_1} \leq m_{\nu_2} \leq m_{\nu_3}$, 
$c_{\alpha i}=\cos{\theta_{\alpha i}}$, $s_{\alpha i}=
\sin{\theta_{\alpha i}}$, 
$\theta_{e2}$, $\theta_{e3}$, $\theta_{\mu 3}$ are three independent
real angles and $\delta$ is a phase responsible for CP violation. \par
We restrict ourselves to the case $\tan^2 \theta_{e 3} \ll 1$, 
in which we get acceptable solutions; in this limit
the experimental bounds [13] can be written as:
%
\be
\cases{\sin^2 2 \theta_{e2} \simeq 3.6 \; 10^{-3} \div 1.2 \; 10^{-2} & \cr
       & \cr
       \sin^2 2 \theta_{\mu 3} \ge 0.49 & \cr       
       & \cr
       \displaystyle {\Delta m^2_{ATM} \over \Delta m^2_{SOL}} = 
       {\Delta m^2_{32} \over \Delta m^2_{21}} \simeq 10^2 \div 2 \; 10^3. & \cr} \ee
The results of the numerical analysis carried out for $M^{II}_{\nu 3}$ are
presented in Fig.1 where we limit ourselves to $0<x_\nu,y_\nu<10$
(from the calculations we have seen that the allowed region of the plane
$(x_\nu,y_\nu)$ is symmetric with respect to the origin). 
The results are weakly dependent on the parameters $x_e$, $y_e$ which we put 
equal to 1 in the matrix $\lambda^E$ needed for the determinations of the mixing 
angles in the lepton sector. \par
We see that all the physical constrains are satisfied in the ranges
%
\be 
\begin{array}{ccc}
6 < x_\nu < 10 & \hbox{and} & -10<x_\nu<-6 \\
5 < y_\nu < 7  & \hbox{and} & -7 <y_\nu<-5
\end{array} 
\ee
in which we obtain the solutions:
%
\be
\cases{\sin^2 2 \theta_{e2} \simeq 3.6 \; 10^{-3} \div 1.2 \; 10^{-2} & \cr
       & \cr
       \sin^2 2 \theta_{\mu 3} \simeq 0.49 \div 0.75 & \cr       
       & \cr
       \tan^2 \theta_{e3} \simeq 10^{-4} \div 10^{-3} & \cr
       & \cr
       \displaystyle {\Delta m^2_{ATM} \over \Delta m^2_{SOL}} = 
       {\Delta m^2_{32} \over \Delta m^2_{21}} \simeq 10^2 \div 3 \; 10^2. & \cr} 
\ee \par
%
%
%
%
%
%
%
%
\begin{figure}
\centering
\includegraphics{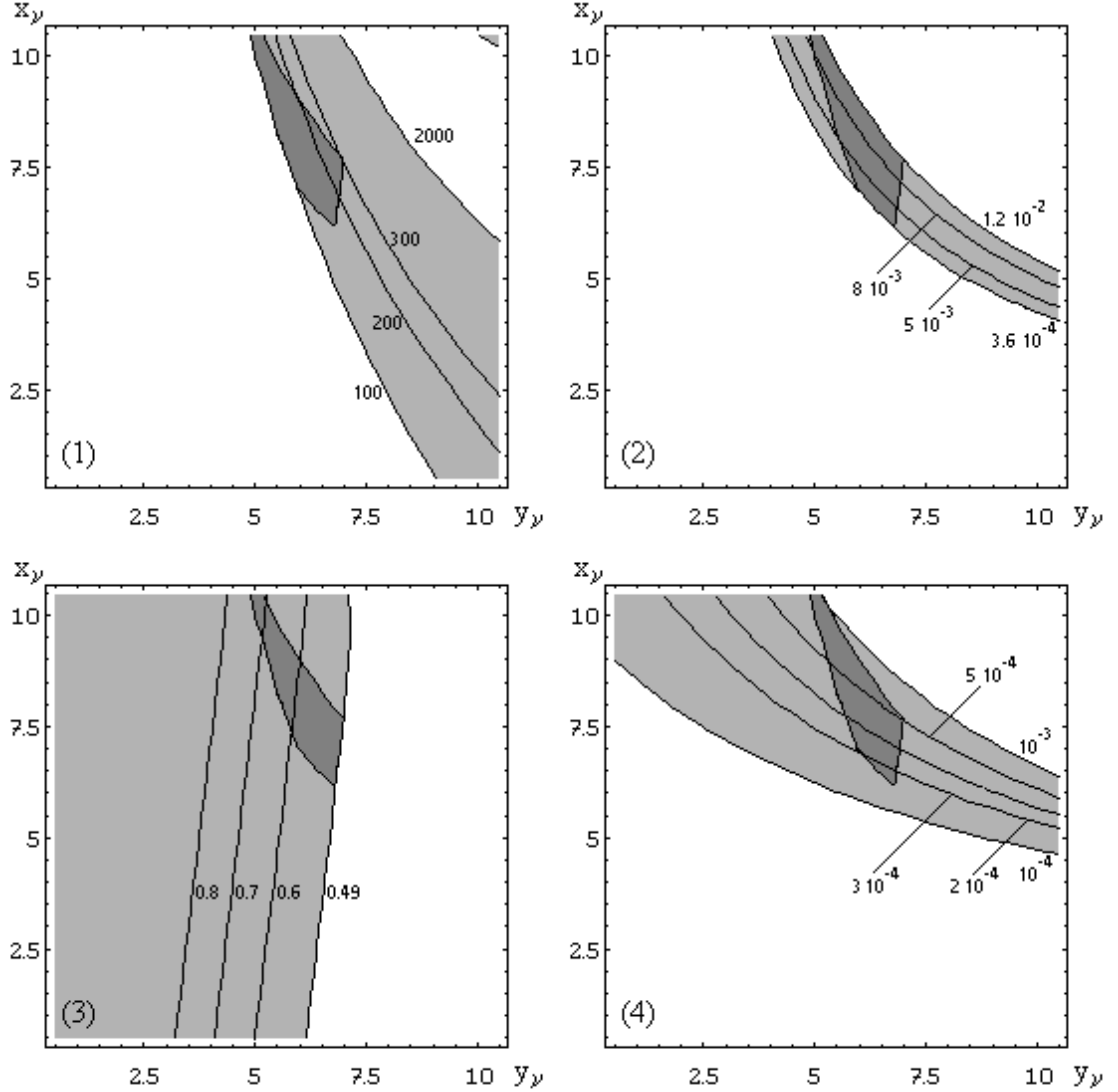}
\caption{
Results of the numerical analysis carried out for $M^{II}_{\nu 3}$ with
$\epsilon = 0.02$, $\epsilon' = 0.004$ and $x_e = y_e =1$.
In the four frames we describe the region of the parameter space allowed
by the experimental data respectively for $\Delta m^2_{ATM} /
\Delta m^2_{SOL}$, $\sin^2 2 \theta_{e2}$, $\sin^2 2 \theta_{\mu 3}$ and
$\tan^2 \theta_{e 3}$. The darker areas are the intersections of the allowed
regions.} 
\end{figure}
%
%
%
%
%
%
%
%
\section{Conclusions}
In this paper we have adopted the seesaw mechanism for the
generation of neutrino masses. With the introduction of a
right-handed neutrino for each family, the masses of the light
neutrinos are given by $M_{\nu} = M_D M_M^{-1} M_D^T$,
where $M_D$ is the Dirac mass matrix, and $M_M$ the Majorana
mass matrix of the heavy right-handed neutrinos.

In the frame of grand-unified models, $M_D$ is related to the
mass matrices of the quarks and charged leptons, while $M_M$
is completely decoupled from the charged sector. As a consequence,
the matrix $M_M$ contains a set of arbitrary parameters, which
makes these models non-predictive.

We have analysed a class of supersymmetric models based on the
$SO(10)\otimes U(2)$ group, where $U(2)$ represents the flavour 
symmetry recently introduced [7,8] to account for the hierarchies
in the mass spectrum of quarks and charged leptons.
The models based on the $SO(10)$ gauge symmetry present the 
advantage, with respect to those based on $SU(5)$ [11], that
all the unknown parameters of the Majorana mass matrix can be
inglobed in a single factor.

The numerical analysis of the neutrino mass matrix shows
that there is a solution, in the allowed region of the
parameter space, which is consistent with the results of the
three-flavour fit [13] of the existing data on solar and
atmospheric neutrinos, and then in favour of the two 
$\nu_e \leftrightarrow \nu_{\mu}$ and $\nu_{\mu}
\leftrightarrow \nu_{\tau}$ oscillation picture.

\end{document}